\documentclass{ws-procs961x669}
\begin{document}

\title{\Large Non-additive \ entropies \  in \  ...  \  gravity ?}
\author{\large Nikos \  Kalogeropoulos}
\address{Weill \ Cornell \ Medicine -  Qatar, \\ Education City, PO Box 24144, Doha, Qatar.\\ 
                                                 E-mail: \  nik2011@qatar-med.cornell.edu} 

\begin{abstract}
We present aspects of entropic functionals  relatively  recently introduced in Physics which are 
non-additive, in the conventional sense of the word, some of which have a power-law functional form.  
We use as an example among them, and to be concrete in this work  the ``Tsallis entropy",  which has, arguably, the simplest functional form. 
We present some of its properties and  speculate about its potential uses in semi-Classical and Quantum Gravity. 
\end{abstract}

\keywords{Non-additive entropy, Tsallis entropy, Semiclassical gravity, Quantum gravity.}

\bodymatter


\section{Introduction and Motivation.}

Two of the contemporary trends in Statistical Physics are the re-examination of its foundations; the reasons and range of its applicability,
and the construction of ``alternative" functionals to that of the Boltzmann/Gibbs/Shannon (BGS) entropy $\mathcal{S}_{BGS}$. 
As is well-known, for a discrete set of outcomes parametrized by $i\in I$ and corresponding probabilities $p_i$.  and with the Boltzmann constant 
being indicated by $k_B$, the BGS entropy is given by 
\begin{equation}
    \mathcal{S}_{BGS}[\{ p_i \} ] \ = \ - k_B \sum_{i\in I} p_i \log p_i 
\end{equation}      
The word ``alternative" above, is actually a misnomer. The proposed entropic functionals have a different domain of applicability 
than the domain of $\mathcal{S}_{BGS}$. One such domain is the case of long-range interactions, such as electromagnetism and gravity, the latter being a 
long-range interaction  in the Newtonian framework. It is not at all obvious, even J.W. Gibbs was a skeptic \cite{T-book},  why $\mathcal{S}_{BGS}$ 
should describe the collective behavior of systems having  such long-range interactions.\\

One way to deal with issues pertaining to the applicability of $\mathcal{S}_{BGS}$, is to construct other entropic 
functionals and develop a corresponding ``thermodynamic formalism" that can be used to make theoretical predictions based on them, which are testable by 
experiments.  Several such functionals have been proposed during the last 30 years, whose form has either been adapted from Information 
Theory, or other disciplines, or have been constructed for the purposes of Statistical Mechanics.\\

 One such prominent example, for which the thermodynamic formalism and properties has been largely worked out is the case of the 
Havrda-Charv\'{a}t/Dar\'{o}czy/Vajda/Lindhard-Nielsen/Cressie-Read/Tsallis \cite{HCT}, henceforth just to be called ``Tsallis entropy" $\mathcal{S}_q$ 
for brevity. Even though the range of applicability of the  Tsallis entropy is still subject to discussion, we choose to focus on $\mathcal{S}_q$  in this work 
due to its simple functional form,  which shares common features with other entropic functionals of power-law form that have been recently proposed. 


\section{Definition and some properties of the ``Tsallis entropy".}

The Tsallis entropy $\mathcal{S}_q$ is a single-parameter family of functionals that are parametrized by $q\in\mathbb{R}$, even though an extension to 
$q\in\mathbb{C}$ has been recently proposed \cite{WW}. Then, with the notation of (1), $\mathcal{S}_q$ is defined as
\begin{equation}
    \mathcal{S}_q [\{ p_i \} ] \ = \ k_B \cdot \frac{1}{q-1} \cdot \left( 1 - \sum_{i\in I} \ p_i^q \right) 
\end{equation}
The extension to continuous systems is not entirely trivial and still subject to some controversy \cite{Abe}. Avoiding all the subtleties, we pretend 
in this work that the naive extension to continuous systems is valid, as most works on this topic tend to assume. Then, let  the
possible set of outcomes form a sample space $\Omega$ endowed with a Borel measure (volume element) $dvol_\Omega$ and a 
probability distribution function (the Radon-Nikodym derivative of the actual measure $\mu$ describing the process, with respect to the 
volume) $\rho: \Omega \rightarrow \mathbb{R}_+$. The Tsallis entropy is then naively defined as 
\begin{equation}
        \mathcal{S}_q [\rho ] \ = \ k_B \cdot \frac{1}{q-1} \cdot \left( 1 - \int_\Omega \ [\rho(x)]^q \ dvol_\Omega \right)
\end{equation}
We choose units so that $k_B=1$ in the sequel. Usually $\Omega$ is chosen to be a finite dimensional Riemannian manifold $(\mathbf{M, g})$ 
with its volume element $dvol_\mathbf{M}$ induced by the metric $\mathbf{g}$, and possibly additional data such as a symplectic structure (when 
$\mathbf{M}$ is the phase space of a system) etc. \\

Some properties of potential interest of the Tsallis entropy are \cite{T-book}
\begin{alphlist}  
   \item The BGS entropy is a member of the family of $\mathcal{S}_q$: 
             \begin{equation}
                   \lim_{q_\rightarrow 1} \  \mathcal{S}_q \ = \ \mathcal{S}_{BGS}
            \end{equation}
   \item Positivity: \  $\mathcal{S}_q \ \geq \ 0.$
   \item Convexity: \ $\mathcal{S}_q$ is concave for $q >0$ and convex for $q<0$. 
   \item Generalized additivity: For two independent occurrences/events $A, B$, namely if the joint probability $p(A\ast B)$ of the combined system $A\ast B$ 
   is expressed in terms of the marginals $p(A)$  and $p(B)$ as \ $p(A \ast B) = p(A)p(B)$, \ then
             \begin{equation}
                  \mathcal{S}_q (A\ast B) \ = \ \mathcal{S}_q(A) + \mathcal{S}_q(B) + (1-q) \mathcal{S}_q(A) \mathcal{S}_q(B) 
             \end{equation}
           This is the most significant difference between $\mathcal{S}_{BGS}$ which is just additive for such independent events, 
           and $\mathcal{S}_q$.  
   \item Sets of axioms distilling the essential properties of $\mathcal{S}_q$ exist, mirroring the better known Shannon and Khintchine axioms 
            defining $\mathcal{S}_{BGS}$.           
\end{alphlist}


\section{Accomplishments and critiques.}

Much like any new proposal, the Tsallis (or any other) of these recently proposed entropic functionals has its strong points but has also its detractors.\\
  
We believe that inasmuch as $\mathcal{S}_q$  provides a statistically different way from $\mathcal{S}_{BGS}$ to encode the important aspects of the 
collective behavior of a system of many degrees of freedom, it is very useful. Naturally, the same can be said about any other entropic functional, especially
Renyi's entropy, which has withstood  very well the test of time and appears in several disciplines. We should notice that there are many aspects and relations of equilibrium 
thermodynamics that can be carried over, usually without or  with minor modifications, to the case of $\mathcal{S}_q$, which among many things, claims 
to also describe some non-equilibrium phenomena \cite{T-book}.  Moreover, there are numerous ongoing 
attempts to apply consequences stemming from $\mathcal{S}_q$ to different disciplines that appear to have only minimal relation to Physics or Information Theory 
\cite{T-book}.   Hence the scope of $\mathcal{S}_q$ may transcend what is traditionally the domain of Physics, in essence following the case of $\mathcal{S}_{BGS}$. \\

Some of the above statements, however, may also be seen as possible shortcomings of $\mathcal{S}_q$: the lack of a concrete set of thermodynamic laws
obeyed by $\mathcal{S}_q$ is somewhat unnerving. In particular, even definitions of basic concepts such as equilibrium,  
temperature etc. exist, but they appear, to us at least, to be somewhat artificial \cite{T-book}. Also:  little is actually known about the nature and ab initio 
calculation of the non-entropic parameter $q$ at both technical and conceptual levels. It is still quite unclear, for instance, what would be the value of $q$ 
when two or more systems having different values of $q$  interact with each other etc, \cite{TT} nonwithstanding. One may have to resort to using the more general 
framework of superstatistics \cite{BC}, rather than remain within the strict confines of the formalism of $\mathcal{S}_q$ but this is still unclear. \\

 The dynamical foundations of $\mathcal{S}_q$ are really not particularly well-understood. Whereas one can point out that ergodic theory and hyperbolic dynamical 
 systems are underlying the use and widespread validity of $\mathcal{S}_{BGS}$, something similar is not really known for $\mathcal{S}_q$, or any of the other recently 
 proposed  entropies. Notice that our comments  throughout this work refer only to Hamiltonian systems having many degrees of freedom. 
 The accepted conjecture is that long (``fat-tailed") spatial and temporal correlations in the underlying system should be responsible for a growth 
 rate of the phase space volume which has a power law rather than exponential form, for the system to be described $\mathcal{S}_q$ \cite{HT}. 
 However very little concrete evidence has been produced to  support or refute such claims. 
 And on top of this, someone can always question why it should be $\mathcal{S}_q$ that describes such systems, and not 
some other entropic functional among the numerous ones that have been constructed which still has a power-law form.


\section{Occasional misunderstandings ...}

It may appear, from the above arguments, that applying the ideas and the formalism of non-additive entropies, such as $\mathcal{S}_q$, to semi-Classical 
or Quantum Gravity may not be a  particularly fruitful endeavor. Indeed, most problems arising in the development of $\mathcal{S}_q$ have become apparent for 
relatively ``simple" particle systems. Far less is known about how to apply such concepts to field theories, and even more so to Quantum Gravity candidates
which have the additional conceptual and technical complications related to diffeomorphism invariance and background-independence.\\ 

It is probably due to the numerous subtleties related to the indefinite signature of the space-time metric, to the diffeomorphism invariance of space-time and to the     
background independence of Quantum Gravity that some misunderstandings have occurred in the part of the literature that attempts to 
apply $\mathcal{S}_q$ to General Relativity and Quantum Gravity. \\

A typical case is that of the holographic nature of black hole entropy. This entropy
depends on the surface area of a space-like section of the event horizon for let's say the Schwarzschild solution. As a result  one might be lead to believe 
that it the entropy of a Schwarzschild black hole should somehow  be related to $\mathcal{S}_q$ for $q\neq 1$ since, after all, $\mathcal{S}_{BGS}$ should be dependent on the 
``volume" of the ``interior" of a black hole rather than its  ``surface" \cite{CT}.  This apparently reasonable, but rather naive picture, is incorrect for multiple reasons. 
First, the volume of the interior of a black hole is not defined. The space-time metric is geodesically incomplete, something that really defines the singularity of a 
space-time \cite{GLG}. Second, even if one disregards the geodesic incompleteness of the metric , 
the volume of the ``interior" of a Schwarzschild solution depends on the choice of the space-like surface \cite{DNM}. We could choose such a  ``volume" to refer to a   
minimal  space-like surface, but this would be an arbitrary and unjustified choice in the classical theory. By contrast, the surface area of any space-like section 
of the event horizon does not depend on the choice of such section and this is what is used in the entropy formulae.  
Third, the Schwarzschild solution is static, so it has a Killing vector field which is space-like in the interior part of the solution. 
Hence, trying to define the volume of the interior of such a solution is like trying to 
determine the volume of a cylinder: such a concept is physically meaningless due to the existence of the translation invariance along the axis of the cylinder 
which is an isometry; the volume of sections which are ``nice" / ``equivariant" under such isometries is what is actually physically relevant.  
  

\section{Possible applications to Semi-classical and Quantum Gravity.}

One cannot help wondering about the possibility of applying $\mathcal{S}_q$ to Gravity. After all, neither black hole entropy nor Quantum Gravity are particularly
well-understood, so it may pay off to find ``unusual" paths toward resolving such problems.
Our current attitude toward this is more favorable than not. 
Some of the reasons are: 
\begin{romanlist}
     \item In the Newtonian framework, gravity is a long-range interaction (in 3 and 4 dimensions). Some studies and corresponding arguments involving long-range 
              interactions, such as in space plasmas, seem to point out, according to their authors at least \cite{Liv}, that the equilibrium distributions that can be derived by applying 
              the Maximum Entropy Principle to $\mathcal{S}_q$, the so-called q-exponentials $e_q(x)$, appear to be a good fit to recently obtained observation data. 
              So, it may be worth considering the possibility of Statistical Mechanics of gravitating systems that relies on $\mathcal{S}_q$.      
     \item It seems that correlations have a significant role to play in understanding the statistical basis of the black hole, and more generally,  the holographic nature 
              of  ``gravitational" entropy. If such correlations turn out to be strong enough to allow the phase space volume to grow in a power-law manner \cite{HT}, 
              but not faster, power-law entropies, such as $\mathcal{S}_q$, may be needed for thermodynamics.
     \item The naive path-integral approach to quantization, which was developed mostly during the 1960s, indicates that gravity is perturbatively (around a Minkowski 
              background, at least) non-renormalizable. The corresponding questions about supergravities \cite{Kal} or string-inspired ``low-energy" actions are still unresolved. 
              So, one is tempted to try to consider path integrals/``partition functions" based on $q$-exponentials and generalized curvature actions, such as      
              \begin{equation}
                     \mathcal{Z}_q \ = \  \int_\mathfrak{M} e_q^{S_q} \  \mathcal{D}_\mathfrak{M}      
              \end{equation}
              where $\mathfrak{M}$ is the space of pairs of (ultimately Euclidean signature) metrics of the background manifold $\mathbf{M}$ and functions 
                  $\phi: \mathbf{M} \rightarrow \mathbb{R}$, with  $\mathcal{D}_\mathfrak{M}$ being a 
              judiciously chosen  measure (such as a $q$-Gaussian) on $\mathfrak{M}$ and a measure on the space of functions $\phi$ 
               determining a conformal deformation of the volume $d\nu = \ e^{-\phi} dvol_\mathbf{M}$. \  
              The action $S_q$, familiar as the action of the Brans-Dicke scalar-tensor theory of gravity in the Jordan frame, is 
              \begin{equation}
                     S_q \ = \ \int_\mathbf{M}  \mathrm{Scal}_q [\mathbf{g}, \phi]  \  e^{-\phi} \  dvol_\mathbf{M}
              \end{equation}
              where the generalized scalar curvature \ $\mathrm{Scal}_q [\mathbf{g}, \phi]$ \  is  given by \cite{Lott}
              \begin{equation}
                      \mathrm{Scal}_q [\mathbf{g}, \phi] \ = \ \mathrm{Scal}[\mathbf{g}] - 2 \cdot \frac{\nabla^2 \phi}{\phi} + \left( 1 - \frac{1}{q} \right) \cdot \frac{\| \nabla\phi \|^2}{\phi^2}   
              \end{equation}
              as is suggested by arguments based on optimal transport \cite{S-LV}. This tacitly assumes some form of long-term ``memory" in the system
              which would most likely lead to an integro-differential, rather than Einstein's, equations in the classical limit, something that would have to be 
              appropriately addressed. 
     \item There are very rich analytic and geometric structures \cite{V-book} associated to $\mathcal{S}_q$ that may be of interest to gravity. 
              In particular, one can provide  synthetic definitions of Ricci 
              curvature which are applicable to spaces with low regularity, such as graphs, Alexandrov spaces etc \cite{V-Tak}.
              It is widely believed that the classical ``smooth" picture of Riemannian manifolds is emergent from something more fundamental and also far less regular.
              In this spirit, structures such as generalized convexity conditions, dimension-dependent Poincar\'{e} and Sobolev inequalities, 
               Wasserstein spaces etc. which are intimately related to  $\mathcal{S}_q$ 
              may be useful in formulating and deriving emergent ``smooth" properties of the underlying space-time, especially in the semi-classical limit.     
     \item The entropy $\mathcal{S}_q$ appears in Perelman's work on the 3-dimensional Poincar\'{e} conjecture \cite{KL}. Since space-times are many times 
              assumed to have the form $\Sigma \times \mathbb{R}$ (if one assumes that the Hamiltonian approach has to be applicable and is willing to consider the 
               possibility of excluding space-time topology change)  where $\Sigma$ is a Cauchy hypersurface which is a Riemannian 3-manifold, 
              $\mathcal{S}_q$ may be relevant in further elucidating aspects of geometrization pertinent to gravity.            
\end{romanlist}


\section{Acknowledgement}

We would like to thank Professor Antonino Marciano, for graciously inviting us to orally present our  
work on aspects of which the present paper relies, in the parallel session that he organized  on ``Emergent and Quantum Gravity" at this Symposium. 




\end{document}